\documentclass[nofootinbib,preprintnumbers,prl,reprint,superscriptaddress]{revtex4-2}
\usepackage{amsmath,amssymb,hyperref,xcolor,cleveref,tikz-cd,mathtools,mleftright,bbm}
\usepackage[final]{microtype}

\begin{document}
\preprint{DMUS-MP-26/03}

	\title{Higher Integrability, Generalized Symmetries, and Partially Topological Field Theories}

	\author{Luigi Alfonsi}
	\email{l.alfonsi@herts.ac.uk}
	\affiliation{Centre for Mathematics and Theoretical Physics, Department of Physics, Astronomy, and Mathematics,\\University of Hertfordshire, Hatfield AL10 9AB, United Kingdom}

	\author{Leron Borsten}
	\email{l.borsten@herts.ac.uk}
	\affiliation{Centre for Mathematics and Theoretical Physics, Department of Physics, Astronomy, and Mathematics,\\University of Hertfordshire, Hatfield AL10 9AB, United Kingdom}

	\author{Mehran \surname{Jalali Farahani}}
	\email{m.jalalifarahani@surrey.ac.uk}
	\affiliation{School of Mathematics and Physics, University of Surrey, Guildford GU2 7XH, United Kingdom}

	\author{Hyungrok Kim}
	\email{h.kim2@herts.ac.uk}
	\affiliation{Centre for Mathematics and Theoretical Physics, Department of Physics, Astronomy, and Mathematics,\\University of Hertfordshire, Hatfield AL10 9AB, United Kingdom}

	\author{Martin Wolf}
	\email{m.wolf@surrey.ac.uk}
	\affiliation{School of Mathematics and Physics, University of Surrey, Guildford GU2 7XH, United Kingdom}

	\author{Charles A. S. Young}
	\email{c.young8@herts.ac.uk}
	\affiliation{Centre for Mathematics and Theoretical Physics, Department of Physics, Astronomy, and Mathematics,\\University of Hertfordshire, Hatfield AL10 9AB, United Kingdom}

	\date{\today}

	\begin{abstract}
We exhibit a classical link between higher Lax integrability and generalized symmetries in a non-Lorentzian higher semi-holomorphic Chern--Simons theory on $M\times\mathbbm CP^1$. The equations of motion determine a spectrally dependent flat higher connection on $M$ whose Laurent coefficients organise an infinite hierarchy of conserved quantities. 
We construct a spectral family of classical \((d_M-2)\)-form symmetry line defects associated with the Lax connection, establish an action-preserving deformation law, and exhibit suitable gauge-invariant probes on which the induced transformations act nontrivially. 
\end{abstract}

	\maketitle

	\section{Introduction and summary}

	\paragraph{Higher integrability.}
	It is difficult to say something interesting  in quantum field theory (QFT). General interacting QFTs are rich in structure but hard. Free field theories are easy but plain in structure. However, in two or fewer spacetime dimensions, \emph{integrable field theories} straddle the boundary between tractability and complexity; they are heavily constrained by an infinite number of symmetries but nevertheless display rich physics. Their study has repeatedly led to profound physical and mathematical insights. 

	Here, we are specifically concerned with \emph{Lax-integrable} field theories, as defined by the presence of an infinite number of conserved charges generated by the structure of a Lax connection. Costello, Witten, and Yamazaki~\cite{Costello:2013zra,Costello:2013sla,Costello:2017dso,Costello:2018gyb,Costello:2019tri} (cf.~the review~\cite{Lacroix:2021iit}, and \cite{Vicedo:2019dej} for the relation to affine Gaudin models \cite{Vicedo:2017cge}) have shown that a plethora of 2d Lax-integrable field theories originate from compactifications of \(d=4\) semi-holomorphic Chern--Simons (CS) theory on Riemann surfaces with defects and boundary conditions. This reformulation through CS gauge theory opens the door to the notion of \emph{higher} integrability. Here, ``higher'' has two related meanings: (i) in higher spacetime dimensions and (ii) in the sense of a higher gauge theory (see e.g.~\cite{Jurco:2019woz,Borsten:2024gox} and the references therein). By replacing the 1-form connection of CS theory by an $\infty$-connection of higher gauge theory, both uplifts are realized at once, yielding various higher Lax-integrable models~\cite{Alvarez:1997ma,Alvarez:1998df,Chen:2024axr,Liniado:2025sin,Benini:2026wys,Benini:2026osk,Alfonsi:2026pth,Chen:2026bct}. In particular, the corresponding Lax $\infty$-connection yields an infinite tower of charges.

	\paragraph{Generalized symmetries.}
A conventional continuous global symmetry in $d$ dimensions 
has
a Noether current whose Hodge dual is a $(d-1)$-form $J$, closed on shell. The corresponding charge $Q(\Sigma)=\int_\Sigma J$ is unchanged under continuous deformations of the $(d-1)$-cycle $\Sigma$, essentially by Stokes theorm. (When $\Sigma$ is a spatial slice, one recovers the ordinary notion of charge conservation.)
In the quantum theory, the corresponding symmetry operator $U_{\alpha}(\Sigma) = \exp(\mathrm i\alpha Q(\Sigma))$ is a topological defect operator of codimension one: its correlation functions are unchanged when $\Sigma$ is smoothly deformed without crossing other insertions \cite[\S2]{Gaiotto:2014kfa}. 

Phrased this way, there is no reason to restrict attention to codimension one. Integrating closed $p$-form currents over $p$-cycles and exponentiating the resulting charge provides a construction (whenever it is well-defined in the quantum theory) of topological defect operators of codimension $(d-p)$; they are called $(d-p-1)$-form symmetries, since their charged objects have dimension $d-p-1$ \cite[\S3]{Gaiotto:2014kfa}. 

It is the topological defect operators themselves that are fundamental. Their fusion, and action on charged objects, encode the generalized symmetry \cite{Gaiotto:2014kfa}; for reviews, see~\cite{Bhardwaj:2023kri,Gomes:2023ahz,Cordova:2022ruw,Brennan:2023mmt,Luo:2023ive,Iqbal:2024pee,McGreevy:2022oyu}.
The standard examples are by no means exhausted by the integrals of local currents. For example, pure $\operatorname{SU}(N)$ gauge theory has a $\mathbb Z_N$ one-form symmetry without an associated Noether current, whose surface generators act on Wilson lines \cite[\S4.2]{Gaiotto:2014kfa}. 
One may also allow non-invertible symmetries. For example, the critical Ising model has a topological Kramers-Wannier duality line with non-group-like fusion \cite{Frohlich:2004ef,Chang:2018iay}.

This broader viewpoint makes holonomies of flat connections, and their higher-gauge analogs, natural candidates for symmetry defects, a suggestion made in \cite[Rem.~2.4]{Schenkel:2024dcd}. 

	\paragraph{Higher integrability and generalized symmetries.}
Here we establish a concrete classical connection between Lax $\infty$-connections and generalized global symmetries. We study a semi-holomorphic higher CS theory on $M\times C$, where $M$ is a real $d_M$-dimensional manifold and $C$ is a punctured Riemann surface, using the model of~\cite{Benini:2026wys}. Its equations of motion produce conserved currents $j$ and $\tilde\jmath$, defining a $(d_M-2)$-form symmetry and a 0-form symmetry, respectively. We make the higher symmetry explicit by constructing spectrally dependent classical line defects, describing their deformation law, and exhibiting gauge-invariant probes on which the induced transformation is nontrivial.

The relevant observables are traced holonomies of the one-form component of the higher Lax connection in this $BF$-type model. Their classical homotopy invariance follows from flatness. Whether the corresponding defects survive quantisation requires a separate anomaly analysis. (In ordinary four-dimensional semi-holomorphic CS theory a framing anomaly obstructs simple closed planar Wilson loops at fixed spectral parameter~\cite[\S3.6 and \S6]{Costello:2017dso}. 
The cotangent-type $BF$ structure of the higher semi-holomorphic CS model in fact suggests a different perturbative behaviour, which we mention below only as a motivation for further study.)

We then isolate the structural ingredients of the example. For a partially topological higher gauge theory on $M\times C$ with a meromorphic higher connection $L$ whose curvature along $M$ vanishes, any well-defined higher holonomy may furnish a homotopy-invariant classical observable. This points towards a more general framework for relating higher Lax connections to generalized symmetries, though one should not assume that classical on-shell flatness alone guarantees the existence of topological quantum defect operators in the quantum theory.

\section{Higher gauge theory}

	\paragraph{Semi-holomorphic Chern--Simons theory.}
	Let us consider a simple, non-Lorentz-invariant theory obtained from the higher semi-holomorphic CS theories introduced in~\cite{Benini:2026wys} that illustrates the appearance of generalized symmetries in higher integrable models. This directly generalizes the considerations of \(d=4\) semi-holomorphic CS theory. We consider the boundary/pole structure in~\cite[Ex.~6.3]{Benini:2026wys} for $M\times C$ with $M$ a $d_M$-dimensional manifold with $d_M\geq2$ and $C\coloneqq\mathbbm CP^1$.

	Consider the differential forms $\Omega_C^\bullet(M\times C)\cong\bigoplus_{k=0}^{d_M}\bigoplus_{q=0}^1\Omega_C^{k;0,q}(M)$ on $M\times C$ where $\Omega_C^{k;0,q}(M)$ denotes the differential $k$-forms on $M$ whose coefficient functions are smooth on $M$ and that take values in the meromorphic $(0,q)$-forms on $C$ with boundary/pole conditions imposed which we specify momentarily. We shall assume that \(\mathrm TM\) admits a global basis $\{E_1,\ldots,E_{d_M}\}$ with $[E_a,E_b]=0$ for all $a,b=1,\dotsc,d_M$. Dually, we then have a global basis $\{e^1,\ldots,e^{d_M}\}$ of $T^*M$ with $\mathrm d_Me^a=0$ for all $a=1,\ldots,d_M$, where $\mathrm d_M$ is the exterior differential on $M$. Hence, any element of $\Omega_C^{k;p,q}(M)$ is of the form $\omega=\tfrac{1}{k!}\omega_{a_1\cdots a_k}\otimes e^{a_1}\wedge\ldots\wedge e^{a_k}$ with $\omega_{a_1\cdots a_k}$ a $(p,q)$-form on $C$ depending smoothly on $M$.

	Next, let $(\mathfrak h,[-,-])$ be a Lie algebra with metric $\langle-,-\rangle$. We take as the gauge \(L_\infty\)-algebra the 2-term $L_\infty$-algebra \(\mathfrak g\coloneqq\mathfrak g_0\oplus\mathfrak g_{2-d_M}\) with $\mathfrak g_0\coloneqq\mathfrak h$ and $\mathfrak g_{2-d_M}\coloneqq\mathfrak h[d_M-2]$ with only the binary bracket being nontrivial and defined by $\mu^{\mathfrak g}_2(x,y)\coloneqq[x,y]$ for all $x,y\in\mathfrak g_0$ or for all $x\in\mathfrak g_0$ and $y\in\mathfrak g_{2-d_M}$.

	Fix distinct $c_1,\ldots,c_{d_M}\in \mathbb R\setminus\{0,1\}$, and consider the meromorphic $(1,0)$-form
	\begin{equation}
		\Omega\coloneqq\frac{\prod_{a=1}^{d_M}(z-c_a)}{z^2(z-1)^{d_M-2}}\,\mathrm dz
	\end{equation}
	on \(C\). The $L_\infty$-algebra underlying our CS theory will now be the tensor product algebra: the underlying vector spaces of fields are\footnote{Technically, for the semi-holomorphic Chern-Simons theory, to be well defined, the fields are valued in the complexified Lie algebra $\mathfrak h^{\mathbb C}$ and subject to standard reality conditions as in \cite[Sec 5.4]{lacroix_4-dimensional_2022}. }
	\begin{subequations}
		\begin{equation}
			\Omega^\bullet_C(M\times C,\mathfrak g)\coloneqq\bigoplus_{p=2-d_M}^{d_M+1}(\Omega^\bullet_C(M\times C,\mathfrak g))_p
		\end{equation}
		with
		\begin{widetext}
			\begin{equation}
				\begin{aligned}
					(\Omega^\bullet_C(M\times C,\mathfrak g))_p&\coloneqq\Omega_C^{p;0,0}(M,\mathfrak h)\oplus\Omega_C^{p+d_M-2;0,0}(M,\mathfrak h[d_M-2])
					\\
					&\kern3cm\oplus\Omega_C^{p-1;0,1}(M,\mathfrak h)\oplus\Omega_C^{p+d_M-3;0,1}(M,\mathfrak h[d_M-2]),
				\end{aligned}
			\end{equation}
		\end{widetext}
	\end{subequations}
	where the coefficient functions $\omega_{a_1\cdots a_k}$ of elements $\omega=\tfrac{1}{k!}\omega_{a_1\cdots a_k}\otimes e^{a_1}\wedge\ldots\wedge e^{a_k}$ in $\Omega^{k;p,q}_C(M,\mathfrak h)$ and $\Omega^{k;p,q}_C(M,\mathfrak h[d_M-2])$ can have simple poles at points $c_{a_1},\ldots,c_{a_k}\in\{c_1,\ldots,c_{d_M}\}$.  The elements of $\Omega^{k;0,0}_C(M,\mathfrak h)$ and $\Omega^{k;0,0}_C(M,\mathfrak h[d_M-2])$ vanish at $z\in\{0,\infty\}$, and, in addition, the elements of $\Omega^{k;0,0}_C(M,\mathfrak h[d_M-2])$ have a zero at \(z=1\) of order at least \(d_M-2\).

	Furthermore, the differential $\mu_1^{\Omega^\bullet_C(M\times C,\mathfrak g)}$ on $\Omega^\bullet_C(M\times C,\mathfrak g)$ is given by $\mu_1^{\Omega^\bullet_C(M\times C,\mathfrak g)}\coloneqq\mathrm d_M+\bar\partial'_C$ where $\bar\partial'_C$ is the regularized antiholomorphic exterior derivative on $C$ defined as in~\cite{Alfonsi:2026pth}; it is the same as \(\bar\partial_C\) except that any Dirac deltas arising from poles are subtracted. The only nontrivial higher product is the binary one $\mu_2^{\Omega^\bullet_C(M\times C,\mathfrak g)}$, which is induced by the wedge product between differential forms and the binary product $\mu_2^{\mathfrak g}$. In addition, $\Omega^\bullet_C(M\times C,\mathfrak g)$ is cyclic with $\langle \alpha,\beta\rangle^{\Omega^\bullet_C(M\times C,\mathfrak g)}\coloneqq\int_{M\times C}\Omega\wedge\langle\alpha,\beta\rangle$, where $\langle-,-\rangle$ on $\mathfrak g$ pairs $\mathfrak g_0$ with $\mathfrak g_{2-d_M}$ via the metric on $\mathfrak h$ (for $d_M\geq3$ this is forced by degree; for $d_M=2$ we assume no $\mathfrak g_0\otimes\mathfrak g_0$ component). The theory is therefore of $BF$ type and  linear in  $\mathfrak h[d_M-2]$-valued fields. Note that the poles of $\alpha$ and $\beta$ are such that the integrand is well-defined.

	Following the lines of~\cite{Alfonsi:2026pth,Benini:2026wys}, it is now not too difficult to establish that the Batalin--Vilkovisky version of the higher CS action (see \cite[Sec 5.2]{jurco_l_infty-algebras_2019})  associated with $\Omega^\bullet_C(M\times C,\mathfrak g)$ is equivalent (or, quasi-isomorphic in homotopy algebra language) to
	\begin{subequations}
		\begin{equation}
			S\coloneqq\int_M\langle\mathrm d_M\gamma,j\rangle+\cdots,
		\end{equation}
		where the ellipsis denotes ghost fields, antifields, etc.,  \(\gamma\in\Omega^{d_M-2}(M,\mathfrak g_{2-d_M})\) and 
		\begin{equation}\label{eq:defOfj}
			j\coloneqq\mathrm e^{-\phi}\mathrm d_{\vec c}\,\mathrm e^\phi
		\end{equation}
		with \(\phi\in\Omega^0(M,\mathfrak g_0)\) and
		\begin{equation}
			\mathrm d_{\vec c}\coloneqq\sum_{a=1}^{d_M}c_ae^aE_a
			\quad\text{with}\quad
			\iota_{E_a}e^b=\delta_a{}^b.
		\end{equation}
	\end{subequations}
	Note that since $\mathrm d_Me^a=0$, we have that $\mathrm d^2_{\vec c}=0$ and $[\mathrm d_{\vec c},\mathrm d_M]=0$.\footnote{The equation $\mathrm d_M(g^{-1}\mathrm d_{\vec c}g)=0$,
with $g=\mathrm e^\phi$, is a special case of the bidifferential
chiral equation in \cite[Sec.~3]{muller-hoissen_frolichernijenhuis_2025}: in the notation of that paper, take $N_1=\mathrm{id}$, $N_2(E_a)=c_aE_a$; note also that 
$g_{\text{there}}= g_{\text{here}}^{-1}$.
}
 The corresponding equations of motion are
	\begin{equation}\label{eq:jeom}
		\begin{aligned}
			0&=-\mathrm d_Mj,
			\\
			0&=-\mathrm d_M\,\mathrm d_{\vec c}~\gamma+[j,\mathrm d_M\gamma].
		\end{aligned}
	\end{equation}
	The higher Lax connection $L$, which descends from $A\in\Omega_C^{1;0,0}(M,\mathfrak h)\oplus \Omega_C^{d_M-1;0,0}(M,\mathfrak h[d_M-2])\subseteq(\Omega^\bullet_C(M\times C,\mathfrak g))_1$, is seen to be
	\begin{subequations}
		\begin{widetext}
			\begin{equation}\label{eq:Lax}
				L\coloneqq-\sum_{b=1}^{d_M}e^b\,\frac{1}{z-c_b}j_b-\sum_{a_i, b=1}^{d_M}\frac{e^b\wedge e^{a_1}\wedge\dotsb\wedge e^{a_{d_M-2}}}{(d_M-1)!}\,\frac{(z-1)^{d_M-2}}{(z-c_b)\prod_{k=1}^{d_M-2}(z-c_{a_k})}\tilde\jmath_{ba_1\cdots a_{d_M-2}},
			\end{equation}
		\end{widetext}
		where $j_b$ are the coefficients of $j$ in~\eqref{eq:defOfj} and $\tilde\jmath_{ba_1\cdots a_{d_M-2}}$ are the coefficients of
		\begin{equation}
			\tilde\jmath\coloneqq\mathrm d_{\vec c}\,\gamma+[j,\gamma],
		\end{equation}
	\end{subequations}
	respectively.

	In the special case $d_M=2$, one can choose \footnote{For $d_M=2$, $z=1$ is no longer a singularity of $\Omega$, so $c_1=1$ is allowed.} $c_1=1$ and $c_2=-1$, and $e^1$ and $e^2$ to be self-dual and anti-self-dual directions, respectively, in Lorentzian signature. Then this theory reduces to the principal chiral model (PCM) whose target space has the Lie algebra $\mathfrak g$. In particular, the equation of motion becomes $-\mathrm d_M{\star J}=0$ for $J=\mathrm e^{-\gamma}\mathrm e^{-\phi}\mathrm d_M \, (\mathrm e^\phi\mathrm e^\gamma)$. Note the identity $\star\mathrm d_M=\mathrm d_{\vec c}$ when acting on functions.

	The currents $j$ and $\tilde\jmath$ are conserved due to the equations of motion, and they define a \((d_M-2)\)-form symmetry and a 0-form symmetry, respectively. The latter generates the right $\mathfrak h$-action. 

	To make the higher charges explicit, consider the Laurent expansion of $L(z)$ about $z=\infty$. Setting $\mathrm d_{(k)}\coloneqq\sum_{a=1}^{d_M}c_a^ke^aE_a$ (so $\mathrm d_{(0)}=\mathrm d_M$, $\mathrm d_{(1)}=\mathrm d_{\vec c}$), which all square to zero and mutually anticommute, we find
	\begin{equation}
		L(z)=-\sum_{k\geq0}z^{-k-1}j^{(k)}+\cdots,
	\end{equation}
	where we have focused on the first term in \eqref{eq:Lax}, since it is the higher symmetry $1$-form current, and defined
	\begin{equation}
		j^{(k)}\coloneqq\mathrm e^{-\phi}\mathrm d_{(k+1)}\mathrm e^{\phi},
	\end{equation}
	with $j^{(0)}=j$. Flatness of $L(z)$ for all $z$ yields the hierarchy
	\begin{equation}
		\mathrm d_Mj^{(k)}=\tfrac12\sum_{p+q=k-1}[j^{(p)},j^{(q)}],
	\end{equation}
	whose $k=0$ member is the equation of motion $\mathrm d_Mj=0$. Write $L_1$ for the first term in~\eqref{eq:Lax}. For a closed curve $\ell\subseteq M$ with base point $x_0\in\ell$, let $\operatorname{Hol}_\ell(z)\coloneqq\mathcal P\mathrm e^{-\int_\ell L_1(z)}$ and expand $\log\operatorname{Hol}_\ell(z)=\sum_{k\geq1}z^{-k}Q_k$. This gives
	\begin{equation}\label{eq:charges}
		\begin{aligned}
			Q_1&=\int_\ell j^{(0)},
			\\
			Q_2&=\int_\ell j^{(1)}+\tfrac12\int_{ s_1<s_2}\mathrm ds_1\,\mathrm ds_2\,[j^{(0)}(s_2),j^{(0)}(s_1)],
		\end{aligned}
	\end{equation}
	and so on, where $\ell^*j^{(k)}=j^{(k)}(s)\,\mathrm ds$ and $s\in[0,1]$ parameterizes $\ell$. This is a direct higher analogue of the non-local charges of~\cite{Brezin:1979am}: here, the current $j=j^{(0)}$ generates the shift symmetry $\gamma\mapsto\gamma+\omega$ ($\mathrm d_M\omega=0$). 
The corresponding 
classical homotopy-invariant observables 
are the traced holonomies $\operatorname{tr}_V\operatorname{Hol}_\ell(z)$ for representation $V$; their $1/z$-expansions contain the invariant combinations of the $Q_k$, such as $\operatorname{tr}(Q_1Q_2)$. For $d_M=2$, the $Q_k$ are the $\mathfrak h$-components of the non-local charges of the PCM with target $\mathrm TH\cong H\ltimes\mathfrak h$. The Casimir has no $\mathfrak h\otimes\mathfrak h$ component, so they Poisson-commute: they form an Abelian ideal, acted on by the charges built from $\tilde\jmath$.

	On the other hand, for $d_M\geq3$, the action $S$ has additional gauge symmetries $\delta\gamma\coloneqq\mathrm d_M\lambda+D_{\vec c}\beta$ with $\lambda,\beta\in\Omega^{d_M-3}(M,\mathfrak h)$ and $D_{\vec c}\coloneqq\mathrm d_{\vec c}+[j,-]$ under which $\phi$ is invariant; $\tilde\jmath=D_{\vec c}\gamma$ is invariant under \(\delta_\beta\) and changes by an exact form under \(\delta_\lambda\) on shell, so that its integral is gauge-invariant.

\paragraph{Classical spectral symmetry defects.}
We now indicate how the conserved quantity $\operatorname{Hol}_\ell(z)$ is associated to $(d_M-2)$-form symmetries of the classical theory, by describing the corresponding symmetry defects and the effect of moving them on classical probe observables.

Fix $z\in\mathbb C\setminus\{0,c_1,\ldots,c_{d_M}\}$ and define rescaling maps $R_k(e^{a_1}\wedge\cdots\wedge e^{a_k}) \coloneqq
\left(\prod_{i=1}^k(z-c_{a_i})^{-1}\right)e^{a_1}\wedge\cdots\wedge e^{a_k}$ for $k=0,1,\ldots,d_M$ and $R_0=\mathrm{id}$. One checks that
$F_{L_1(z)}=-zR_2(\mathrm d_Mj)$ holds identically, and moreover that the action $S = \int_M \langle \mathrm d_M \gamma,j\rangle$ can be re-written in the form\footnote{We assume $\int_{\partial M}\langle\gamma\wedge j\rangle=0$; otherwise this boundary term must be added to the right-hand side.}
\begin{equation}S[g,B_z]=\int_M\langle B_z, F_{L_1(z)}\rangle,\end{equation}
where we define $B_z\coloneqq\frac{(-1)^{d_M}\prod_{a=1}^{d_M}(z-c_a)}{z}R_{d_M-2}\gamma$, a $(d_M-2)$-form depending on $\gamma$, and where $F_{L_1(z)}\coloneqq\mathrm d_M L_1(z)+\tfrac12[L_1(z),L_1(z)]$ is the (off-shell) curvature of the Lax connection.
For this classical construction we keep $g=\mathrm e^\phi \in H$ smooth and allow $\gamma$, hence $B_z$, and the multiplier gauge parameters to be currents. 
Classical defect lines associated to holonomy operators (such as our $\operatorname{Hol}_\ell(z)$) are discussed in e.g.~\cite[\S3.1]{Alekseev:2015hda}. Such defects carry a degree of freedom, namely a field $h\colon S^1 \to H$. One picks a $\lambda \in \mathfrak h$ labelling a coadjoint orbit of $H$,
and the defect action is then
\begin{equation} I_{\ell,\lambda}[h;g] \coloneqq \int_{S^1}\langle\lambda, h^{-1}\mathrm d_{S^1}h+h^{-1}\iota^*(L_1(z))h\rangle\end{equation}
where $\iota : S^1 \to M$ is the embedding map with image $\ell=\iota(S^1)$.
Now let $X:[0,1]_t \times S^1_s \to M$ describe a smooth motion of the defect line, from $X(0,s) = \iota_0(s)$ to $X(1,s) = \iota_1(s)$, with $\ell_i=\iota_i(S^1)$ for $i=0,1$.
Under this motion, the internal degree of freedom transforms by parallel transport for the pull-back of $L_1(z)$. 
Let $K_X$ be the $\mathfrak h$-valued $(d_M-2)$-current defined by the demand that
\( \int_M \langle K_X, \omega \rangle \coloneqq \int_{[0,1]\times S^1} \langle h\lambda h^{-1}, X^*\omega\rangle
\) 
for $\omega\in \Omega^2_c(M,\mathfrak h)$ any compactly-supported smooth test form. 
One then has the (off-shell) identity
\begin{equation}
 S[g,B_z]+I_{\ell_0}[h_0;g] =  S[g,B_z-K_X]+I_{\ell_1}[h_1;g]
\end{equation}
which identifies the coupled variational problems. As one expects of a higher symmetry, it says that one may move the defect, transforming its field as one goes, at the cost of modifying the bulk fields along the support of the swept-out 2-chain. 

Thus the combined defect motion and field transformation preserve the coupled action off shell. Homotopy coherence modulo gauge transformations, though, holds only on shell. Namely, \emph{on solutions of the coupled bulk--line equations}, 
homotopic motions with fixed initial and final curves induce, for $d_M\geq3$, the same transformation modulo (distributional) gauge transformations \(\gamma\mapsto\gamma+\mathrm d_M\lambda+D_{\vec c}\beta\), with \(g\) unchanged and \(\lambda,\beta\) compactly-supported \(\mathfrak h\)-valued \((d_M-3)\)-currents.\footnote{In two dimensions the multiplier shifts associated to homotopic motions agree exactly on shell; no gauge parameter of degree $-1$ is required.} 

At fixed $z$, one may fuse cooriented symmetry defects, retaining independent internal fields, with the understanding that the products give rise to actions of the form $I_{\mathrm{fused}}=I_{\ell,\lambda_1}[h_1;g]+I_{\ell,\lambda_2}[h_2;g]$ which generally do not correspond to single coadjoint orbits. Under this product fusion, defects with positive-dimensional coadjoint orbits are non-invertible.

To test nontriviality of these symmetries, it suffices to find a gauge-invariant classical observable -- defined at least on a locus preserved by the gauge transformations and the defect motion -- and show that the symmetries act non-trivially on it.
For $d_M\geq3$, suppose that $M$ contains a closed coordinate $(d_M-2)$-torus $\Sigma$, tangent to the $E_a$ with $a\neq r,s$. We orient $\Sigma$ compatibly with the orientation of $M$ and the ordered normal frame $(E_r,E_s)$. Now consider field configurations with constant $g|_\Sigma$.
Both the gauge transformations $\gamma\mapsto\gamma+\mathrm d_M\lambda+D_{\vec c}\beta$ and the defect motion preserve this locus, since they leave $g$ unchanged. On this locus, the quantity $P_\xi\coloneqq\int_\Sigma\langle\xi,\gamma\rangle$, for constant $\xi\in\mathfrak h$, is gauge invariant whenever the required pullbacks of the fields and gauge parameters exist: the tangential components of $j$ vanish, and the remaining tangential derivative terms integrate to zero. If the surface swept out by the moving line defect intersects $\Sigma$ transversely, the result is a change
\begin{equation} \Delta P_\xi= \frac{(-1)^{d_M+1}z}{(z-c_r)(z-c_s)}
         \sum_{u\in X^{-1}(\Sigma)}\epsilon_u\langle\xi,q_X(u)\rangle,
\end{equation}
in the value of $P_\xi$, where
 $\epsilon_u$ is the intersection sign relative to $\mathrm dt\wedge\mathrm ds$ and the ordered normal frame $(E_r,E_s)$, and $q_X=h\lambda h^{-1}$.
For $g=1$, constant nonzero $q_X=q$, a single crossing and $\langle\xi,q\rangle\neq0$, the response is nonzero and therefore cannot arise from $\gamma\mapsto\gamma+\mathrm d_M\lambda+D_{\vec c}\beta$.
(In two dimensions there are no such gauge shifts, and $P_\xi(p)=\langle\xi,\gamma(p)\rangle$ is a point observable wherever its value exists.)

The question of whether these symmetry defects survive in the quantum theory requires further analysis. Suppose, however, that the present model admits an anomaly free quantization preserving the defect-deformation identity. The classical identity would then yield Ward identities for renormalized insertions. For example, suppose there exists a disorder defect \(\mathcal D_\xi(\Sigma)\) whose defining boundary conditions require the holonomy around a small, positively oriented loop linking \(\Sigma\) once to be conjugate to
$$ M_\xi(z)=\exp\!\left(\frac{z\,\xi}{(z-c_r)(z-c_s)}\right),$$
and that this prescription and the usual Wilson-character evaluation survive quantization. Then, for a loop \(\ell\) homotopic to the \(m\)-fold meridian in the complement of all insertions, one would obtain
 \begin{equation}\label{eq:eigenvalue}
 \big\langle U_\ell^V(z)\,\mathcal D_\xi(\Sigma)\,\mathcal O\big\rangle
 =
 \chi_V\!\left(M_\xi(z)^m\right)
 \big\langle\mathcal D_\xi(\Sigma)\,\mathcal O\big\rangle,
 \end{equation}
where $U_\ell^V(z)=\operatorname{tr}_V\operatorname{Hol}_\ell(z)$, $\chi_V$ is the character of $V$, $\mathcal O$ denotes other renormalized insertions avoided by the homotopy, and \(m\) counts the signed traversals of the small loop around $\Sigma$.

Regarding potential anomalies, recall that, in particular,  4d semi-holomorphic Chern-Simons theory suffers a framing anomaly that obstructs simple closed planar Wilson loops at fixed spectral parameter \cite[\S3.6~and~\S6]{Costello:2017dso}.  Superficially, such framing-constraints are in tension with the topological nature of $U_\ell^V(z)$ at fixed $z$. However, the $BF$ structure of the parent theory in the present case means that the only propagator connects the $\mathfrak h$- and $\mathfrak h[d_M-2]$-valued components, and every vertex contains exactly one of the latter. Operators built from $L_1$ alone, such as $U^V_\ell(z)$, therefore receive no self-interaction corrections. This suggests that, in contrast with $d=4$ semi-holomorphic CS theory, there is no framing anomaly for such operators and no $\hbar$-shift of the spectral parameter, and~\eqref{eq:eigenvalue} is exact.  On the other hand, operators involving $\tilde\jmath$ can self-contract, and framing effects may  arise. The possible framing-like constraints they may impose represent interesting avenues for further work.

(For $d_M=2$, the cotangent Yangian and its double provide a natural setting in which to investigate these quantum line-operator structures
\cite{Abedin:2024cqs,abedin2025doubleyangianfactorizationqkzequation}.)

	\paragraph{General flat higher connections.}
One may expect to generalize the construction above, as well as those of~\cite{Chen:2024axr,Benini:2026wys}. Let $M$ be a $d_M$-dimensional real manifold and \(C\) a complex manifold of \(d_C\) complex dimensions. Let us consider a partially topological theory on \(M\times C\) with a gauge potential
	\begin{equation}
		L\in\left(\Omega^\bullet_C(M,\mathfrak g)\right)_1\coloneqq\bigoplus_{k+i=1}\Omega_C^k(M,\mathfrak g_i),
	\end{equation}
	valued in an \(L_\infty\)-algebra \(\mathfrak g=\bigoplus_{i=3-d^{\mathbbm C}}^0\mathfrak g_i\) with $d^{\mathbbm C}\coloneqq d_M+d_C$, where $\Omega_C^k(M,\mathfrak g_i)$ is the space of $\mathfrak g_i$-valued differential $k$-forms on \(M\) whose coefficient functions are smooth on $M$ and meromorphic on $C$ obeying suitable boundary/pole conditions. Suppose that the equations of motion imply that \(L\) is flat along $M$ in the higher-gauge-theory sense:
	\begin{equation}\label{eq:eom}
		F_M(L)\coloneqq\mathrm d_ML+\sum_{k\geq1}\frac{1}{k!}\mu_k(L,L,\ldots,L)=0.
	\end{equation}
	Here, the $\mu_k$ are the higher $L_\infty$ brackets generated by the tensor product of the $C_\infty$-algebra of $\Omega^\bullet_C(M)$ with the $L_\infty$-algebra $\mathfrak{g}$.  The action could be a simple \(BF\)-type action \(\int_{M\times C} \omega \wedge \langle B\wedge F_M(L)\rangle\), where $\omega$ is a meromorphic top form along $C$ and \(B\)  is a $(0,d_C)$-form along $C$ with  legs  along \(M\) as required.

	Flatness~\eqref{eq:eom} is the local condition required for higher parallel transport, but it does not by itself guarantee that a gauge-invariant observable exists for every closed submanifold. When a representation or higher representation, the relevant global data, and a well-defined notion of higher holonomy are available, one may form spectrally dependent holonomy observables along suitable cycles~\cite{chen73,chen75,chen77,igusa2009iteratedintegralssuperconnections}. Their deformation invariance must then be checked in the given theory, together with boundary conditions, singularities, and possible quantum anomalies. Under these additional hypotheses, such observables are natural candidates for generalized-symmetry defects in the sense of~\cite{Gaiotto:2014kfa}.
	Any solution to~\eqref{eq:eom} may be deformed by a  transformation \(g\colon M\times C\to G\), where $G$ is the $\infty$-group corresponding to $\mathfrak{g}$. These are  pseudo-gauge transformations in the sense that they may not respect the relevant boundary conditions; if \(g\) has nontrivial behavior near the singularities and defects in \(C\), this will generically generate physically inequivalent new solutions. This generalizes the notion of dressing transformations in two-dimensional integrable theories, which correspond to pseudo-gauge transformations of the Lax connection \(L\) in \(d=4\) semi-holomorphic CS theory.

\section{Conclusion}
We have given a concrete classical example in which a higher Lax connection organises generalized symmetries. In the non-Lorentzian higher semi-holomorphic CS model, the equations of motion imply the flatness of a spectrally dependent higher connection on $M$. Its Laurent expansion yields a hierarchy of conserved quantities, while the leading current defines a $(d_M-2)$-form symmetry. We constructed the corresponding classical spectral line defects, described their motion by an off-shell identity in the coupled bulk--defect system, and exhibited gauge-invariant probes on which this motion acts nontrivially. Assuming the absence of suitable framing anomalies, this would enhance to generalized symmetries of the quantum theory.

The results of this paper suggest that a partially topological higher gauge theory with a flat meromorphic higher connection is a natural setting in which to search for higher-integrable structures and generalized symmetries.
It would be interesting to test this framework in cohomological theories in, for instance, twistorial formulations of self-dual Yang--Mills theory \cite{Dolan:1982dc,Chau:1982mj,Popov:1995qb,Popov:1998pc,Wolf:2004hp}, self-dual gravity \cite{Penrose:1976js,Atiyah:1978wi,Manin:1988ds,Wolf:2007tx,Lechtenfeld:1999ik,Hoegner:2012sq,Dunajski:2026dfe}, and perhaps even the formulation of full super-Yang--Mills theory on ambitwistor space~\cite{Witten:1978xx,Isenberg:1978kk,Movshev:2004ub,Mason:2005kn,Borsten:2024cfx} for which hidden classical symmetries are known to exist \cite{Popov:2006qu}.

Finally, the spectral parameter need not be intrinsically complex. Away from singular points, a real smooth parameter also admits a local Taylor expansion, although such an expansion does not capture nonperturbative dependence and singular behaviour is less controlled. This leaves open the possibility of analogous constructions in partially topological theories with real, rather than holomorphic, non-topological directions~\cite{Alfonsi:2025kmj}.

\textbf{AI usage statement.}
The authors used large language models as an aid in developing and checking some of the proofs and calculations, and in improving the clarity of the exposition. They revised the final text and take full responsibility for the contents of the manuscript
. 
\textbf{Data statement.}
No additional research data were generated or used beyond those presented and cited in this manuscript.

\textbf{Licence statement.}
For the purpose of open access, the authors have applied a Creative Commons Attribution (CC-BY) license to any author-accepted manuscript version arising.
\bibliography{biblio}

\end{document}